# Differentiating anomalous and topological Hall effects using first-order reversal curve measurements

Gregory M. Stephen[1*], Ryan T. Van Haren[1], Vinay Sharma[1], Lixuan Tai [2,3], Bingqian Dai [3], Hang Chi [4], Kang L. Wang [3], Aubrey T. Hanbicki[1], Adam L. Friedman[1*]

[1]*Laboratory for Physical Sciences, College Park, Maryland 20740 USA*

[2]*RIKEN, Center for Emergent Matter Science (CEMS), Wakō, Japan*

[3]*Department of Electrical and Computer Engineering, University of California, Los Angeles, California 90095 USA*

[4]*Department of Physics, University of Ottawa, Ottawa, Ontario KIN 6N5, Canada*

*Corresponding Authors: gstephen@lps.umd.edu, afriedman@lps.umd.edu

**Abstract**

Next generation magnetic memories rely on novel magnetic phases for information storage. Novel spin textures such as skyrmions provide one possible avenue forward due to their topological protection and controllability via electric fields. However, the common signature of these spin textures, the topological Hall effect (THE), can be mimicked by other trivial effects. Competing anomalous Hall effect (AHE) components can produce a peak in the Hall voltage similar to that of the THE, making clear identification of the THE difficult. By applying the first-order reversal curve (FORC) technique to the Hall effect in candidate topological Hall systems we can clearly distinguish between the THE and AHE. This technique allows for quantitative investigation of the THE and AHE in magnetic materials and heterostructures with topologically non-trivial spin textures. We demonstrate the technique and apply it to several examples.

Recent research claiming the observation of the topological Hall effect (THE) is abundant for a variety of novel materials and heterostructures including perovskites, metallic multilayers, Heusler compounds, and intermetallics [1–4]. However, the expected signature of THE, a peak in the Hall voltage normally seen on top of an Anomalous Hall effect (AHE) hysteresis, can also result from overlapping AHEs with different hysteretic behaviors. [5–7]. The Hall effect is essentially a measure of various scattering mechanisms and is therefore a powerful tool for understanding many intrinsic material properties such as carrier density, type, and mobility [8,9]. These properties are crucial to understanding how different materials can be utilized for practical electronics including next-generation memory, logic, and sensors [10–13]. Because the THE and AHE arise from scattering mechanisms that both occur in magnetic systems, it is vital to develop techniques that can unambiguously differentiate between them.

The current standard for qualitatively differentiating between the THE and AHE is by analyzing minor loop magnetic hysteresis behavior [6,7]. While this method is reasonably successful at differentiating THE and AHE in systems with little noise or complexity, results from samples with multiple co-existing phases are still ambiguous [5–7]. A solution to definitively and quantitatively differentiate between the two effects is the first-order reversal curve (FORC) technique. FORC is widely employed on bulk magnetic materials and geological samples to identify different magnetic phases by measuring the remagnetization characteristics of the sample [14–16]. The magnetic moment is measured returning to saturation for various demagnetizing (reversal) field strengths. Plotting the second partial derivative of the magnetic moment as a function of both applied field and reversal field creates a fingerprint of magnetic phases. FORC has been used to study the magnetoresistance and switching fields of magnetic tunnel junctions important for memory applications [17], but its use is not widespread outside of geologic magnetometry and permanent magnets. Intriguingly, the technique can be generalized to any system with a hysteresis [17,18]. While the measurement itself can be time consuming, requiring many partial hysteresis loops with enough resolution to get an accurate derivative, it provides detailed information about the hysteretic behavior that would otherwise be impossible to acquire.

In this letter, we investigate the FORC technique to identify suspected topological Hall samples. First, we present the general model of the FORC technique as applied to these types of systems. We simulate results to demonstrate the differences between systems with trivial and topological spin texture. Then, we apply the experimental method using DC fields and currents to generate hysteretic magnetoresistance curves to identify the AHE components in $Mn_3Sn/Bi_{85}Sb_{15}$ films, as well as reanalyze published data [6,7] on suspected examples of AHE/THE and ambiguous AHE films. We confirm that our method provides result that are consistent with their conclusions and demonstrate that the technique can clearly separate the Hall effects and distinguish a topological film from a trivial one.

The physical origins of the AHE and THE are quite similar, and a unified theory was presented by Verma, Addison, and Randeria [19]. Both are transverse voltages induced by electrons passing through a material with ordered spins. The spins create an emergent electromagnetic field due to the Berry curvature, which causes transverse scattering of the carriers, and thus a measured Hall voltage. The difference between the AHE and THE is the origin of the Berry curvature. The AHE is governed by the Berry curvature in momentum-space, while the THE stems from the Berry curvature in position-space [20,21]. Both effects exist in the absence of an external magnetic field. With an applied magnetic field, they develop according to the figure of merit for the Berry curvature. For AHE, the figure of merit is the magnetization along the applied magnetic field, causing $\rho_{xy}^{AHE} \propto m_{\parallel}(H)$. For the THE, the figure of merit is the skyrmion density. As skyrmions generally only exist within a range of magnetic field values, this manifests as a peak in $\rho_{xy}$ [22,23]. The two effects are shown pictorially in **Fig. 1(a, b)**.

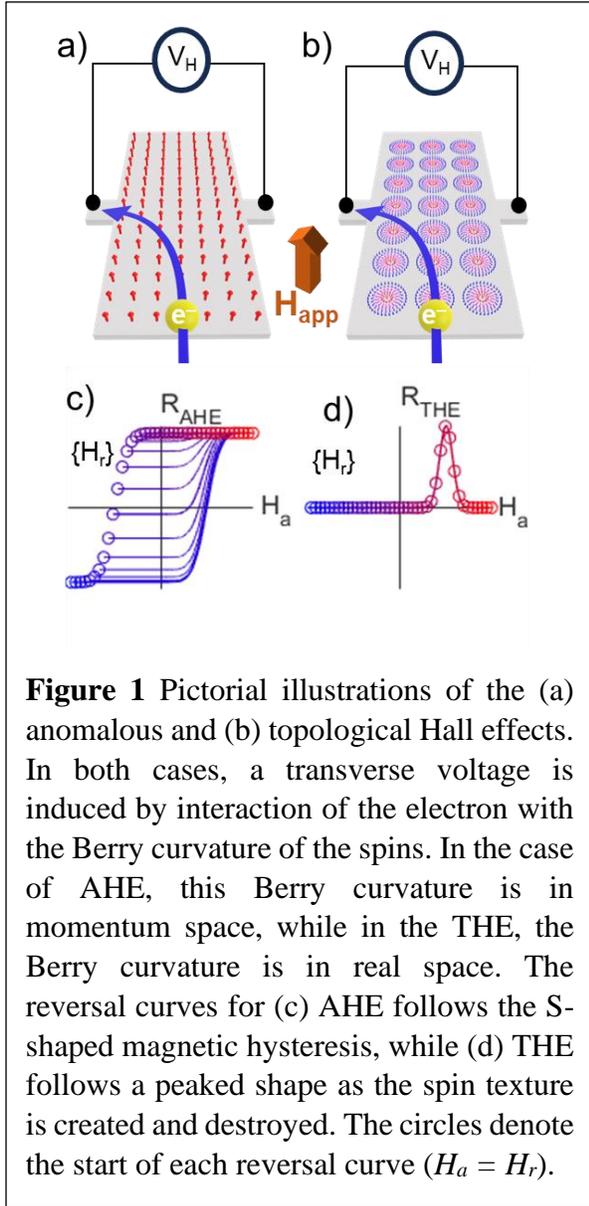

**Figure 1** Pictorial illustrations of the (a) anomalous and (b) topological Hall effects. In both cases, a transverse voltage is induced by interaction of the electron with the Berry curvature of the spins. In the case of AHE, this Berry curvature is in momentum space, while in the THE, the Berry curvature is in real space. The reversal curves for (c) AHE follows the S-shaped magnetic hysteresis, while (d) THE follows a peaked shape as the spin texture is created and destroyed. The circles denote the start of each reversal curve ($H_a = H_r$).

Experimentally, the FORC technique is most often applied to magnetization versus magnetic field. However, the same method can be applied to any hysteretic behavior [14,17,18]. Here, we apply this method to field-dependent resistance, $R(H_a, H_r)$, where $H_a$ is the applied magnetic field and $H_r$ is the reversal field. In measuring $R$ using FORC, the field is first increased to saturate the magnetization then decreased to $H_r$. Afterwards, $R$ is measured as $H_a$ is increased back to saturation. This is repeated for a range of $H_r$ values, producing a family of curves as seen in **Fig. 1(c)**. The FORC distribution is defined as

$$\rho_{FORC}(H_a, H_r) = \frac{\partial^2 R}{\partial H_a \partial H_r}. \quad (1)$$

By calculating this mixed partial derivative, any components that do not depend on both $H_a$ and $H_r$ are removed. To make the results more intuitive, we can rotate the axes from applied and reversal fields $(H_a, H_r)$ to the more physically relevant quantities of coercive and interaction fields $(H_c, H_u)$, where $H_c = (H_a - H_r)/2$ and $H_u = (H_a + H_r)/2$. In magnetometry or the AHE, the interaction field $H_u$ is the exchange bias. This rotation allows us to relate the results to the intrinsic coercive and interaction fields within the material rather than just the applied fields.

The positions of peaks in $\rho_{FORC}(H_c, H_u)$ are directly correlated to the effective fields in the sample. Because THE appears as a peak in the measured Hall voltage rather than an S-shaped saturation [5,7], its signature in the FORC distribution appears as a grouping of positive and negative peaks. Some noise could be eliminated by directly measuring $\partial R/\partial H_a$ via a DC + AC magnetic field and a lock-in amplifier, thereby requiring only one calculated derivative. Nonetheless, sufficiently small steps in $H_a$ and $H_r$ provide equivalent resolution. While a large THE component may be apparent in the raw measured reversal curves, the FORC method can distinguish smaller THE components that could be easily overwhelmed by a large AHE component. Fundamentally, the FORC technique separates functions of the applied magnetic field based on their hysteretic character.

To numerically simulate AHE, we use a simplified Stoner-Wohlfarth model assuming that each AHE component is composed of a sum of square hysteresis loops [24]. Each loop has a coercive $H_{c,i}$ and interaction field $Hu_{,i}$. In total, one AHE component is modeled as

$$R(H_a, H_r) = \frac{R_0}{N}\sum_{i=0}^{N} \begin{cases} 1 & if\ H_{u,i} - H_{c,i} < H_r \\ sgn(H_a - H_{c,i} - H_{u,i}) & if\ H_{u,i} - H_{c,i} \geq H_r \end{cases}, (2)$$

$$\text{where } sgn(x) = \begin{cases} -1 & if\ x < 0 \\ 0 & if\ x = 0 \\ 1 & if\ x < 0 \end{cases}.$$

The coercive and interaction fields are assumed to be normal distributions around an average of $H_c$ and $H_u$, respectively. The number square hysteresis loops, $N$, is chosen to be large enough to create a smooth hysteresis loop. **Fig. 1(c)** shows the reversal curves for a simple AHE system calculated with **Eq. 2**. The model ignores any angular distribution of the magnetic moments or any relationships between coercive and exchange fields that would allow peaks in the FORC distribution to be non-elliptical or to rotate.

To numerically simulate THE, we assume it is a simple peaked function without hysteresis as described in experimental literature [25,26]. For simplicity, we assume a Gaussian THE peak shape, though the results can be generalized to any peak shape. As this function is not inherently hysteretic, the entire family of reversal curves overlap, as seen in **Fig. 1(d)**.

**Fig. 2** shows a comparison of the simulated reversal curves and FORC distributions for several different systems. In **Fig. 2(a,b)**, we show the simulations for a system with two different AHE components with coercive fields of 5 kOe and 0.5 kOe (AHE + AHE). In **Fig. 2(c,d)**. we simulate a system with an AHE component with $H_c$ = 5 kOe and a THE component peaked at a field of 4 kOe. In the FORC distributions, yellow(blue) corresponds to positive(negative) values of $\rho(H_c, H_u)$. In both cases, a full hysteresis loop would show a peak on top of a conventional hysteresis. However, the reversal curves, **Fig. 2(a,c)**, and FORC distributions, **Fig. 2(b,d)**, are radically different. The boundary of the AHE+AHE reversal curves traces out the full hysteresis loop and the FORC distribution shows two bright spots corresponding to high $H_c$ (yellow) and low $H_c$ (blue) coercive field components with positive and negative AHE coefficients, respectively. The AHE+THE signal modeled in **Fig. 2(c)** has a peak in all reversal curves. In the FORC distribution (**Fig. 2(d)**) the same high $H_c$ peak appears as in (**Fig. 2(b)**), however a pair of peaks now appear as adjacent positive and negative peaks along the vertical $H_u$ axis (indicated by the red arrow in the figure). In this example, the pair is centered at 4 kOe on the $H_u$ axis, which corresponds to the center of the THE peak. Thus, a pair of peaks separated on the $H_u$ axis in FORC distributions is the signature of a THE component.

While the difference is clear for these two examples where the THE and AHE components are the same order of magnitude, the technique also works when the THE is significantly smaller than the AHE component. **Figure 2(e,f)** shows the same AHE and THE components as **Fig. 2(c)**, but with the THE amplitude reduced by a factor of 100. The THE component is no longer immediately visible in the reversal curves (**Fig. 2(e)**); however, it is still clear in the FORC distribution (red

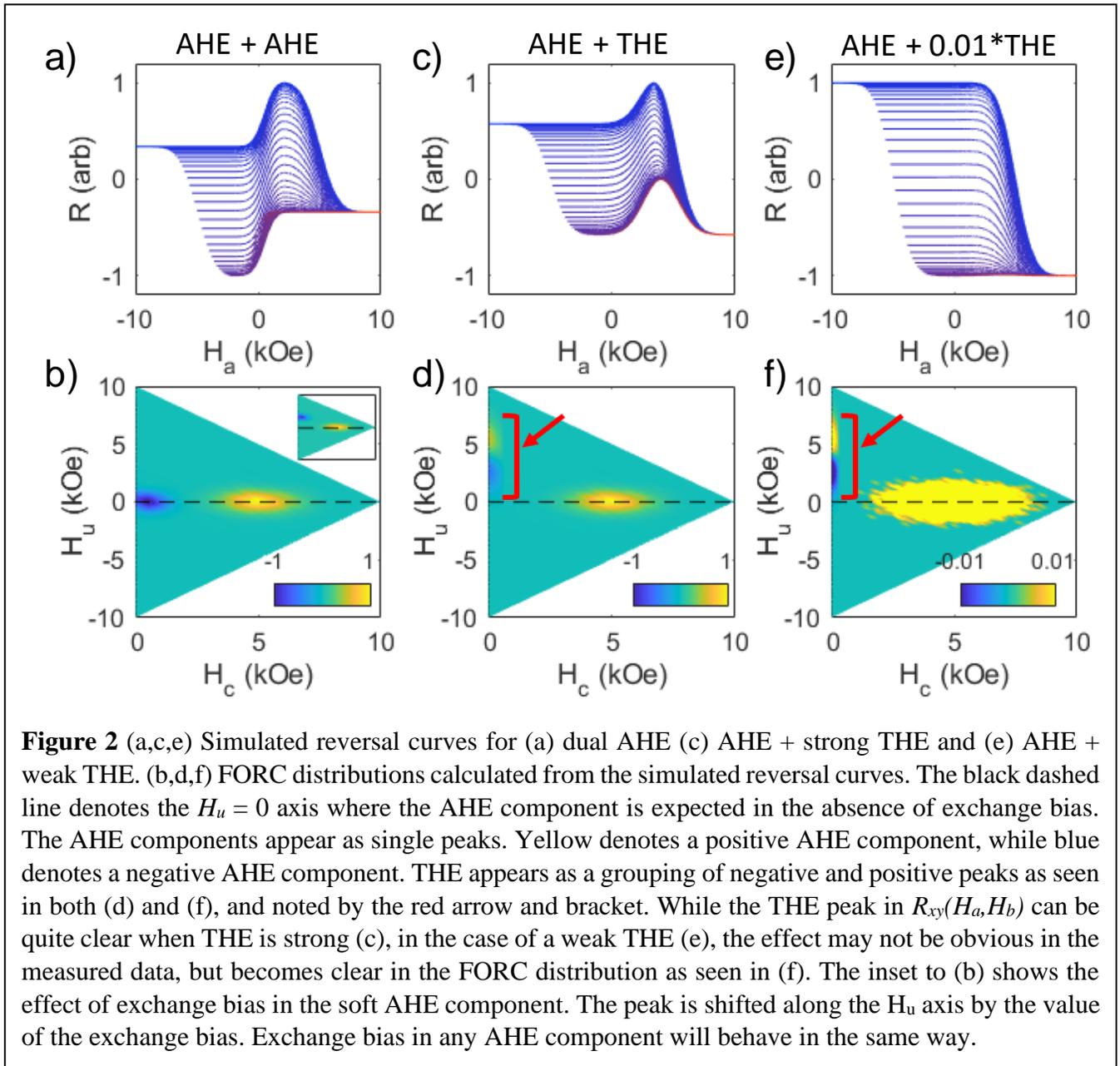

**Figure 2** (a,c,e) Simulated reversal curves for (a) dual AHE (c) AHE + strong THE and (e) AHE + weak THE. (b,d,f) FORC distributions calculated from the simulated reversal curves. The black dashed line denotes the $H_u = 0$ axis where the AHE component is expected in the absence of exchange bias. The AHE components appear as single peaks. Yellow denotes a positive AHE component, while blue denotes a negative AHE component. THE appears as a grouping of negative and positive peaks as seen in both (d) and (f), and noted by the red arrow and bracket. While the THE peak in $R_{xy}(H_a,H_b)$ can be quite clear when THE is strong (c), in the case of a weak THE (e), the effect may not be obvious in the measured data, but becomes clear in the FORC distribution as seen in (f). The inset to (b) shows the effect of exchange bias in the soft AHE component. The peak is shifted along the $H_u$ axis by the value of the exchange bias. Exchange bias in any AHE component will behave in the same way.

arrow - **Fig. 2(f)**). In the case of real data, although the THE component may be near the noise floor of the measurement, the FORC analysis can still distinguish it from the noise. The FORC technique itself does not limit the sensitivity.

The simplified Stoner-Wohlfarth model used in **Eq. 2**. excludes contributions from angular distributions of magnetic domains and correlations between $H_c$ and $H_u$. These interactions would allow for non-elliptical peaks in the FORC distribution. Additionally, the lack of hysteresis in the THE forces the THE peaks to be along the $H_u$ axis. In the case of hysteretic THE, the pair of peaks would be shifted along $H_c$ by the degree of hysteresis, but remain a coupled pair of positive and negative peaks.

To demonstrate the usage of the FORC technique with real data, we applied it to our measurements of a $Mn_3Sn/Bi_{85}Sb_{15}$ heterostructure grown *in situ* using sputtering, as well as data previously published by Tai *et al.* [7] and Chi *et al.* [6].

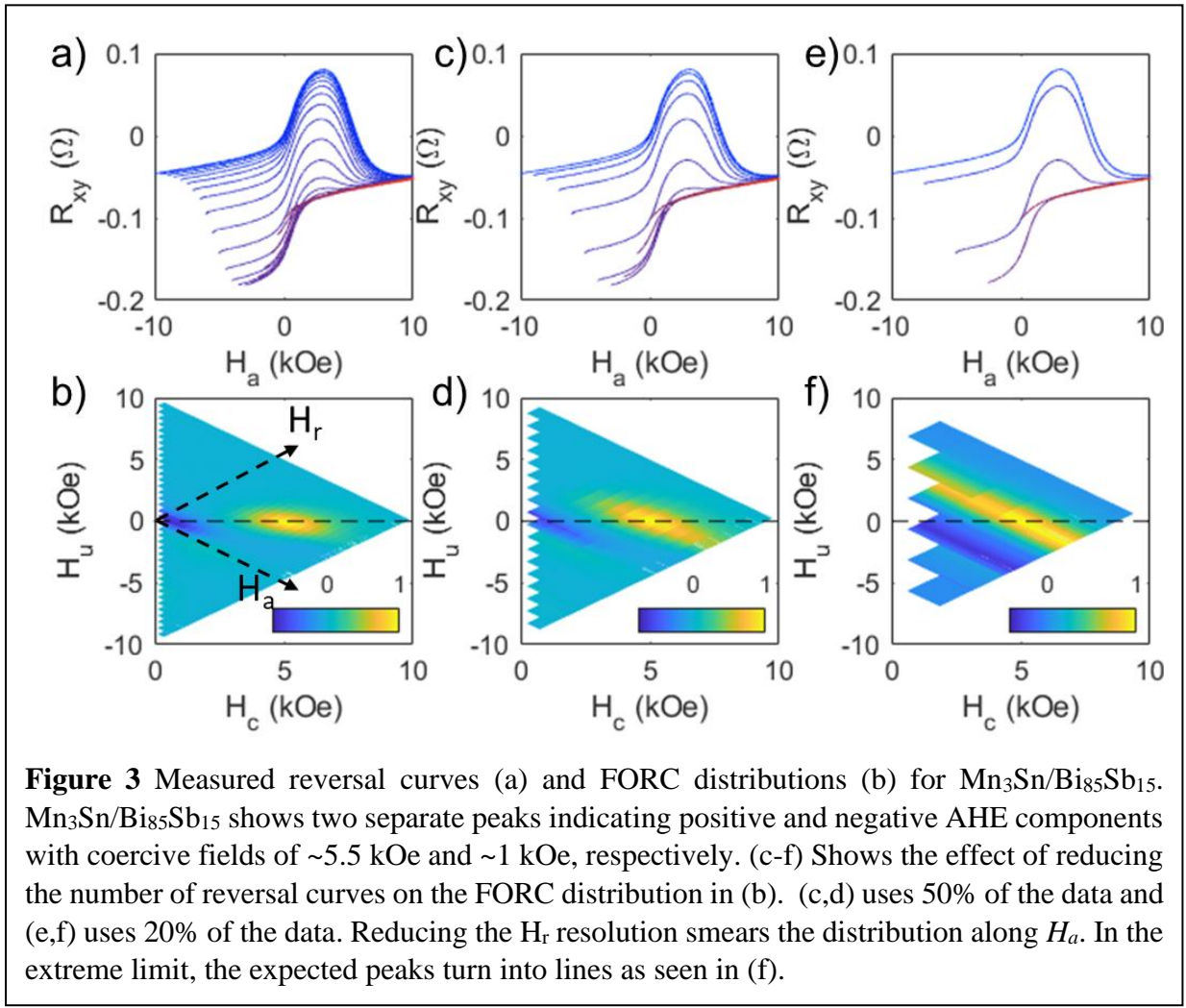

**Figure 3** Measured reversal curves (a) and FORC distributions (b) for $Mn_3Sn/Bi_{85}Sb_{15}$. $Mn_3Sn/Bi_{85}Sb_{15}$ shows two separate peaks indicating positive and negative AHE components with coercive fields of ~5.5 kOe and ~1 kOe, respectively. (c-f) Shows the effect of reducing the number of reversal curves on the FORC distribution in (b). (c,d) uses 50% of the data and (e,f) uses 20% of the data. Reducing the $H_r$ resolution smears the distribution along $H_a$. In the extreme limit, the expected peaks turn into lines as seen in (f).

**Figure 3** shows the measured reversal curves and FORC distribution for $Mn_3Sn/Bi_{85}Sb_{15}$. $Mn_3Sn$ is a non-collinear antiferromagnet with a strong AHE, and there have been reports of a THE contribution [27–29]. The curves in **Fig. 3(a)** show the characteristic peaks of the AHE, and an overall slope due to the ordinary Hall effect. The FORC distribution in **Fig. 3(b)** clearly shows two peaks: a positive peak (yellow) corresponding to a high $H_c$, positive $R_{AHE}$ component, and a negative peak (blue) corresponding to a low $H_c$, negative $R_{AHE}$ component. The FORC technique allows us to unequivocally demonstrate that our $Mn_3Sn/Bi_{85}Sb_{15}$ samples do not have a contribution from the THE, but rather two opposing AHE components. **Figure 3(b)** looks qualitatively like **Fig. 2(b)** as they both have approximately the same distributions of $H_c$ and $H_u$ for the two components.

As minor loop data is not normally taken for THE samples, it is difficult to determine whether published data is genuinely showing the THE or just multiple AHE components. When minor loop data is available, there are often insufficient sweeps for the required $H_r$ resolution. In these cases, the data tend to smear along $H_a$, or towards the lower righthand corner in the FORC distribution. **Figure 3(a,b)** uses 40 reversal curves to achieve sufficient resolution in the FORC distribution. **Fig. 3(c-f)** show the effect of sparse data on the calculated FORC distribution. **Figure 3(c,d)** show the results using 50% of the measured reversal curves, while **Fig (e,f)** show 20%. As the number of $H_r$ curves is reduced, the peak in the FORC distribution smears along the $H_a$ direction, which is towards the lower right corner of the plot. Understanding how the amount of data used to calculate $\rho(H_c, H_u)$ allows us to interpret FORC analysis of existing literature data where only a few minor loops are available.

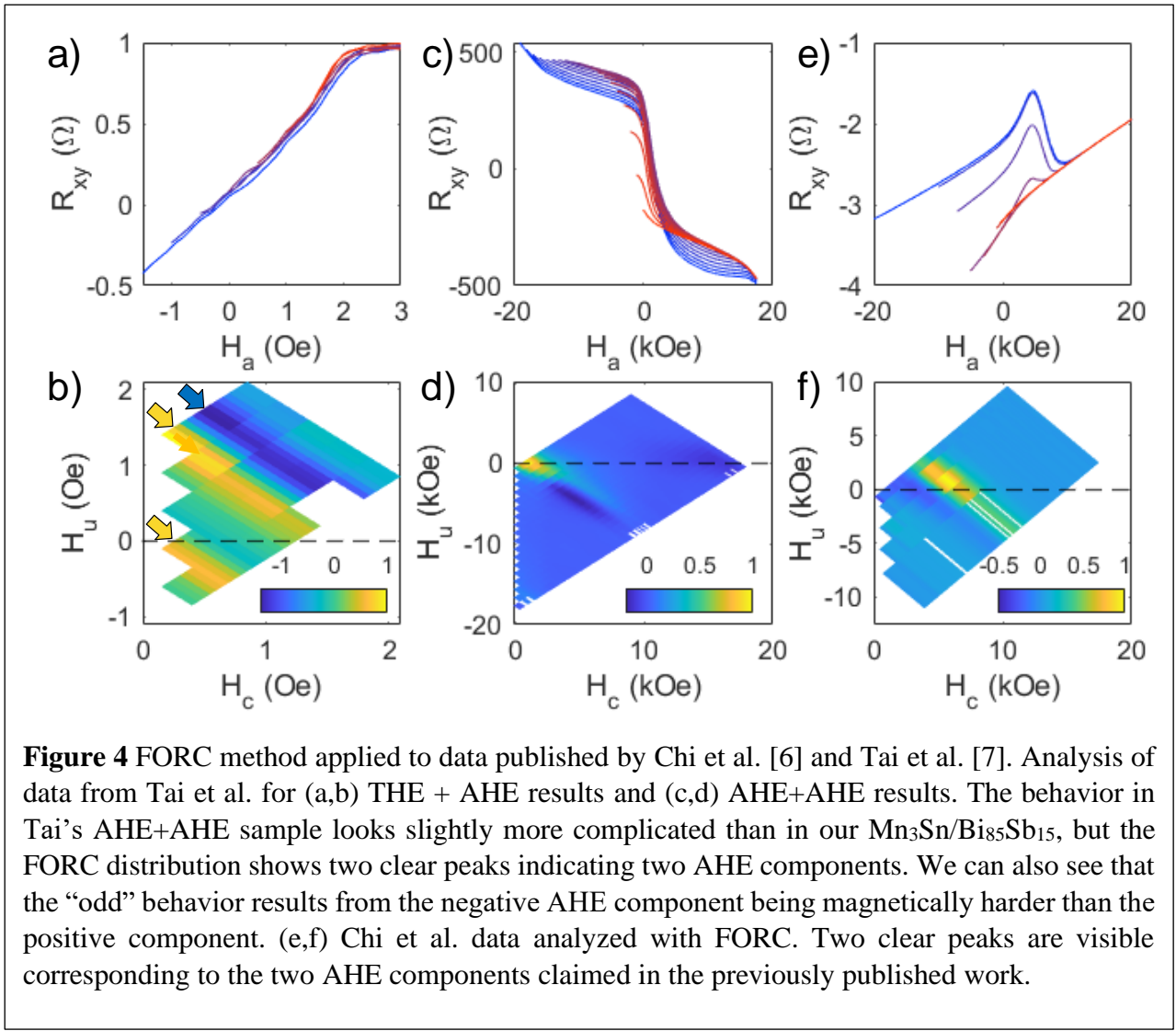

**Figure 4** FORC method applied to data published by Chi et al. [6] and Tai et al. [7]. Analysis of data from Tai et al. for (a,b) THE + AHE results and (c,d) AHE+AHE results. The behavior in Tai's AHE+AHE sample looks slightly more complicated than in our $Mn_3Sn/Bi_{85}Sb_{15}$, but the FORC distribution shows two clear peaks indicating two AHE components. We can also see that the "odd" behavior results from the negative AHE component being magnetically harder than the positive component. (e,f) Chi et al. data analyzed with FORC. Two clear peaks are visible corresponding to the two AHE components claimed in the previously published work.

Tai *et al*. used minor loops to study the THE in Ta/CoFeB/Ir/MgO/Al$_2$O$_3$ stacks and two-component AHE in MnBi$_2$Te$_4$ [7]. Chi et al. studied what appeared to be the THE in Cr$_2$Te$_3$ and determined it was actually two-component AHE due to the distribution of magnetic domains and interfacial strain [6]. In all cases, the FORC technique supports the conclusions made in the respective papers. **Figure 4** shows the measured reversal curves and the corresponding FORC distribution for these previously published data. **Figure 4(a,b)** show the reversal curves and FORC distribution for the Ta/CoFeB/Ir/MgO/Al$_2$O$_3$ stack. Despite the low resolution due to the limited number of reversal curves taken, there is a clear positive/negative pair of peaks centered at *($H_c$, $H_u$)* = (0.75 Oe, 1.0 Oe), plus a bright positive peak at ( -0.5 Oe, -0.5 Oe). These peaks are consistent with the coexistence of the THE and AHE, as was concluded in the published results based on the minor loops tracing back on one another. Again, the smearing of the peaks is due to the limited number of reversal curves available, as we demonstrated would be the case for sparse data in our simulations in **Fig. 3(d,f)**.

Reversal curves and FORC distributions from MnBi$_2$Te$_4$ are shown in **Fig. 4(c,d)**. The FORC distribution shows two distinct peaks corresponding to two AHE components. The curves in **Fig. 4(c)** look qualitatively different from the data in **Fig. 3(b)** because, unlike the case of Mn$_3$Sn/Bi$_{85}$Sb$_{15}$, for MnBi$_2$Te$_4$ the high $H_c$ component is negative (blue) and the low $H_c$ component is positive (yellow). On close inspection, there may be an additional broad negative peak near the end of the $H_c$ axis, though more data would be required to confirm. Given the complex magnetic interactions between layers in MnBi$_2$Te$_4$, additional AHE components would not be surprising [30]. **Figure 4(e,f)** show the data from Chi, et al. along with the FORC results. Again, the data are smeared along $H_a$ due to the limited number of $H_r$ sweeps, but there are still two clear peaks indicating dual AHE behavior.

In conclusion, FORC leverages and expands upon existing qualitative techniques to quantitatively separate the AHE and THE components. This removes analytic ambiguity in THE analysis using minor magnetic hysteresis loops. In the FORC distribution, the AHE components appear as single peaks with positions corresponding to their respective coercivities and exchange biases, while the THE components appear as pairs of peaks centered along the $H_u$-axis at the THE peak positions. While the basic technique described here requires a large amount of data that is not typically gathered to get definitive resolution, widely available image processing techniques could likely allow for clear results with less measured data. Some of these methods are already utilized in tools and software developed for FORC magnetometry. In sum, this technique simplifies the analysis of the THE results allowing for more quantitative analysis.

**Acknowledgements**

The authors from LPS gratefully acknowledge critical assistance from LPS support staff including G. Latini, J. Wood, R. Brun, P. Davis, and D. Crouse.

**Competing financial interests**

The authors declare no competing financial interests.